# Ultra Buck DC/DC Converter for Electric Vehicles

Luoqi Chen
School of Electronics, Electrical
Engineering and Computer Science
(EEECS), Queen's University Belfast
Belfast, BT9 5AH, UK
lchen16@qub.ac.uk

Ahmad Elkhateb
School of Electronics, Electrical
Engineering and Computer Science
(EEECS), Queen's University Belfast
Belfast, BT9 5AH, UK
A.Elkhateb@qub.ac.uk

Nikolaos Athanasopoulos
School of Electronics, Electrical
Engineering and Computer Science
(EEECS), Queen's University Belfast
Belfast, BT9 5AH, UK
N.Athanasopoulos@qub.ac.uk

*Abstract*—A critical challenge in power conversion in electric vehicles is the efficient use of DC-DC buck converters that need to provide 12-V supply for load systems from 400/800-V batteries. This paper presents a literature review on the development of DC-DC buck converters. Moreover, one novel four-phase interleaved step-down topology is selected for simulation and hardware experiments. Based on the four-phase interleaved structure, an extended-phase topology is proposed, which has a higher voltage conversion ratio. Control techniques are also applied to it. Theoretical analyses and simulation results are provided to verify the improved converter. A 400V-to-12V and 150W output power hardware prototype is implemented to verify its performance

## I. Introduction

The objective of this paper is to identify the long-term challenges in the design of power converters and conditioning systems for electric vehicles. Since conventional converters cannot meet the requirements for stepping down the high voltage in electric vehicles, a current challenge is to develop an ultra-high buck converter that converts 400/800V to 12V, that can be used by the load. The main challenges that have been studied in this work are continuous input current, ripple, semiconductors losses, and the conversion ratio.

In conventional DC-DC buck converters, active switches are directly connected in series with the input power supply so that the input current is discontinuous. Due to the discontinuous input current, electromagnetic interference (EMI) can be generated. A converter employing one-cycle control (OCC) with a continuous current is presented in [1]. Many solutions have been proposed in the literature to avoid discontinuous input current. Three different buck converters are introduced in [2]. In the proposed converters, continuous input/output current can be achieved by adding series input or output inductance. The topology proposed in [3] also uses interleaved structure, and it can achieve the continuous current with extended duty cycle. For most of the newly proposed structures, a continuous input current can be achieved, but the challenge is providing continuous current with the minimum number of components and higher efficiency. In [4], with the implement of isolated transformers, the converter can operate in CCM while the input current can flow in two interleaved phase in both modes.

The generation of ripple increases circuit losses reduces efficiency, and even shortens device life. Transformers are always used for output ripple cancellation. By applying an autotransformer in the converter, the ripple voltage can be reduced to 0.35%, with an overall efficiency of 96% [5]. Since transformers make circuits complicated and costly, more researchers tend to use other components to replace transformers. In [6], the presented n-stage SC buck converters can eliminate the output voltage ripple with the addition of a large number of switches. A topology is proposed in [7], which is based on the interleaved discharging (ID) approach, can have reduced output voltage ripple by implementing a two-stage switched-capacitor and a wide range of the duty cycle for actual operation.

The main drawback of the traditional buck converter is that the output current ripple can only be cancelled by adding high inductors. The higher inductors can cause higher losses and cost, which lowers the overall efficiency. The emergence of interleaved buck converters (IBC) has led to new options for solving this problem [8-10]. Due to the phase shift between interleaved phases, the current stress of topologies can be cut down. The superposition of the interleaved phase currents at the output also allows the output ripple to be eliminated [11]. Nevertheless, the challenge of reducing the inductor size and ripple remains in conventional IBC. The work in [12] presents an interleaved buck topology with improved conversion ratio. Therein, the output current ripple can be considerably low, with a tiny inductor.

Many ripple cancellation techniques using a tapped inductor have been proposed [4, 13-16]. On the basis of these, with the addition of coupled inductors, the proposed converter in [16] can achieve ripple removal and a higher conversion ratio. Series capacitors and transformers are applied to this converter, which can also result in the low voltage rating components. By combining a half-bridge converter and a forward converter, a novel interleaved converter is formed, with small output current ripple [4]. The ripple current caused by these two primary converters can be cancelled out by each other in load side.

Switching and conduction losses are main contributors to power loss in power converters. In conventional or interleaved buck converters, the voltage pressures for switches or diodes are forced by the input voltage. Only the costly high voltage rated switches should be applied to in these converters. The challenge of reducing voltage stress of semiconductors in topologies and improve the efficiency is also one of the critical aspects in current DC-DC converters development. [17] introduces a family of converters with a single transistor and two diodes. A modified converter presented in [18] avoids many drawbacks that still exist in [17], such as switching voltage stresses. With the addition of an extra voltage-divider circuit, the added blocking capacitors can reduce the voltage stresses of switches [19]. In [20], the voltages of two switches and diodes are cut down by two-thirds of the input voltage during the switching on or off period.

Nowadays, higher frequency is desirable because it can enable a reduced size of transformers and passive components, such as the sizes of output filters, hence the converter size and weight will be reduced. However, higher switching frequency may bring more power loss in devices. In addition, as switching frequency increases, switching losses increases since the duty cycle becomes small.

Soft switching technology also brings a new direction to reduce the switching losses. The idea is to reduce switching losses by enforcing the voltage or current to be zero while the switch is turning on or turning off. This is called zero-current switching (ZCS) and zero-voltage switching (ZVS). In resonant converters, switching only occurs when the voltage, the current, or both, is zero. Thus, by avoiding instantinous switching of voltage and current, switching loss can be eliminated. Different methods are used to achieve ZVS or ZCS in converters [21-24].

For the purpose of reducing voltage stress of semiconductors, the switching frequency is increased in [25]. This paper also demonstrates that power switches can both operate under ZVS and ZCS to improve efficiency. The work [22] presents two buck converters with zero-current switching. All the semiconductors can achieve soft-switched only with the addition of small inductors while these inductors can delay the conduction current. However, a large capacitor is needed, which can bring high losses.

By adding a coupled inductor, ZVS can be achieved [26]. In [27], an interleaved step-down converter with low switching loss is proposed. Although zero-voltage-transition (ZVT) is achieved in the presented topology, switching loss still exist during the turn off period of the auxiliary switch. In [28], an IBC which only operates at discontinuous conduction mode (DCM) is proposed. Although the switches can be turned on under ZCS, they still need to be turned off by hard-switching. Therefore, the active switches and diodes in presented converter still need to withstand large current pressure. In [23], a more efficient and convenient method was proposed. By combining two buck converters and an added circuit, all the semiconductor elements in the proposed converter can operate under zero voltage switching condition.

The current industrial development requires for the development of step down converters with high conversion ratio in order to handle the new requirments of high voltage batteries with minimum number of conversion stages and higher efficiency. But for the conventional buck converter, the only method to have a low output voltage is to adjust the duty cycle to a very small number, which has many drawbacks. The addition of coupled inductors, switched-capacitors, cascading combination and switched inductors are common to extend the duty cycle of converters. A simple buck topology that reduces the voltage gain is proposed in [29]. The conversion ratio can be reduced to half of the conventional step converter with the same duty cycle, but the efficiency is not high. For the purpose of achieving a high voltage ratio with high efficiency, [30] proposes a novel buck converter with the addition of switched capacitor and switched inductor, which can extend the duty cycle.

Coupling inductors and transformers are also often applied to extend the voltage conversion ratio. Many solutions are introduced in [16, 31-36] which can adjust the turns ratio to extend the duty cycle. Concerning the topology presented in [31], although the conversion ratio is improved, the circuit becomes more complicated due to the use of two transformers. In [16], due to the four phases interleaved structure and the zero current switching turn-on condition, the current stresses of the switches are cut down significantly. With extremely low output ripple, the presented converter can achieve higher efficiency. The voltage gain can also be adjusted by changing the coefficient of the transformers. Nevertheless, the use of two coupling transformers makes the circuit more complicated, and consequently more expensive. By adding two series inductors and a clamp capacitor in the topology in [32], the high voltage conversion and soft-switching operation can be achieved. The other drawback of using coupled inductors in the topology, is the presence of leakage inductance. Although the transformers can bring many disadvantages, their effect on increasing the conversion ratio makes them still used.

To facilitate reductions in both cost and size, many step-down/step-up topologies with or without transformers are presented [19-20, 27, 39-51]. [36] presents an ultra-buck converter which contains a series clamp capacitor. Voltage conversion ratio can be improved through the improvement of topology. However, the presence of the discontinuous output current and high voltage stress for switches still needs to be solved.

In [27], an IBC with improved voltage conversion ratio is proposed. Although the voltages of active switches are bucked, the switches still suffer from hard switching and diodes face reverse recovery problems. The interleaved buck converter proposed in [19], has one fourth of normal IBC voltage conversion ratio. Neverthtless, most of the switches should still suffer high voltage, which causes more conduction losses. In [37], the proposed ultra buck converter has lower voltage stress for transistors with improved conversion ratio. But the inductor current in two phases is unbalanced if the variation of duty cycles in the two switches becomes large.

With addition of two energy-transferring capacitors, the converter can achieve an automatic current balance in every phase in [38]. However, the active switches are still subject to high voltage pressure. Comparedto the topology in [19], the voltage gain is similar, even though the topology in [38] uses more components.

In [39], an interleaved four-phase step-down topology is presented. In the presented topology, lower switches voltage can be achieved without the addition of transformers. Since the capacitors can share the voltage, the step-down ratio can be increased. In this case, three switches suffer high voltage which may lead to low efficiency. Another similar four-phase interleaved DC-DC buck converter is proposed in [20]. Extended duty cycle is achieved by the addition of one extra inductor and four extra capacitors. Current can also be shared equally in four interleaved modules.

All above aspects can be considered as long-term challenges in the design of step-down converters. In general, the novel buck converters proposed so far are improved not only in one aspect, but usually with several improvements at the same time. The development of microelectronic devices pushes converters to smaller sizes for use on smaller devices. Efficient converters are always needed and the maturity of soft switching technology provides a new option for converters to reduce switching losses. Many advanced control techniques that can be used for buck converters are proposed to achieve automotive power conversion and conditioning systems.

The rest of this paper is organized as follows. Section II presents presents analysis and discussion of the proposed buck topology. Simulation and hardware experiment of the selected topology are obtained. Improvements to the selected four-phase converter are presented in Section III, and theoretical analyses and simulation results are provided to verify the improved converter. Finally, the conclusions are drawn in Section IV.

## II. ANALYSIS OF THE PROPOSED TOPOLOGY

In order to design a DC/DC buck converter with high voltage conversion ratio (800V-to-12V), several topologies with improved voltage conversion ratio have been simulated first, while one topology is selected.

### A. Operation of selected topology

The selected convert is proposed in [20]. In Fig. 3, it is a four phase interleaved step down converter. The proposed converter consists of four active switches $S_1$–$S_4$, four freewheeling diodes $D_1$–$D_4$, five inductors, including four filter inductors and one output inductor, and five capacitors, including two input capacitors, two blocking capacitors and one for the output filter. It can be observed that the four active switches control the operation status of the four phases respectively.

In an ideal situation, the proposed converter consists of eight operation modes in one switching period. In the first mode, $S_1$ is on and all diodes are conducting expect $D_1$. At this time, all other switches are not working. $L_1$ is charging through two paths, while $L_2 - L_8$ are discharging.

$$V_{L1}(t) = V_{in} - V_{CB1} - V_{C2} = V_{C1} - V_{CB1} - V_{Lo}(t) - V_o \quad (1)$$
$$V_{Lo}(t) = V_{C1} + V_{C2} - V_{in} - V_o \quad (2)$$
$$V_{L2}(t) = V_{L3}(t) = V_{L4}(t) = -V_o - V_{Lo}(t) \quad (3)$$
$$I_{Lo}(t) = I_{L1}(t) + I_{L2}(t) + I_{L3}(t) + I_{L4}(t) - I_{in}(t) \quad (4)$$

In second mode, $S_1$ is turned off and all switches are off. Since the inductor current cannot become zero immediately, the current in inductor $L_1$ still flows in two paths. All diodes are conducting and all inductors ($L_1 - L_4$) are discharging.

$$V_{L1}(t) = V_{L2}(t) = V_{L3}(t) = V_{L4}(t) = -V_o - V_{Lo}(t)$$
$$= V_{in} - V_{C1} - V_{C2} \quad (5)$$
$$V_{Lo}(t) = V_{C1} + V_{C2} - V_{in} - V_o \quad (6)$$

In every quarter of a switching cycle, only one switch is closed, and all switches are then turned opened. The four switches in the converter cause 16 different working states, but in all the operation modes, the converter is also working similarly as mentioned above. For example, when $S_1$ and $S_2$ are turned on, both $L_1$ and $L_2$ are charging while the other two inductors are discharging. By applying the volt-second balance (VSB) for all inductors ($L_1 - L_4$), equations can be derived as below:

$$(V_{C1} - V_{CB1} - V_o) \cdot D - V_o \cdot (1 - D) = 0 \quad (7)$$
$$(V_{CB1} - V_o) \cdot D - V_o \cdot (1 - D) = 0 \quad (8)$$
$$(V_{C2} - V_{CB2} - V_o) \cdot D - V_o \cdot (1 - D) = 0 \quad (9)$$
$$(V_{CB2} - V_o) \cdot D - V_o \cdot (1 - D) = 0 \quad (10)$$
$$V_{CB1} = V_{CB2} = \frac{V_{C1}}{2} \quad (11)$$
$$V_{C1} = V_{C2} \quad (12)$$

Then the voltage gain equation can be calculated as

$$\frac{V_o}{V_{in}} = \frac{D}{4 - D} \quad (13)$$

### B. Simulation and hardware experiment

The proposed converter not only has an extended duty cycle, but also low voltage stress for diodes and switches, which leads to losses reduction. In order to verify whether the converter proposed in the paper can meet the expectations and whether the design analysis is reasonable, the simulation of this converter is carried out in Simulink. The topology of this simulation is to buck 400V input voltage to 24V output, while output power is 500W. With the 30kHz switching frequency, the duty cycles of all the switches are set to 23.5%. And the parameters of components used are also stated in Table 1. $L_1 - L_4$ are designed for continuous inductor current mode (CICM) condition. Fig. 4 shows the switching signal of all switches ($S_1$-$S_4$).

Table 1. Components' parameters

| Components | Specification |
|---|---|
| $V_{in}$ | 400V |
| $V_{out}$ | 24V |
| $P_{out}$ | 500W |
| Duty cycle | 21.5% |
| Switching frequency | 30 kHz |
| $L_1 - L_4$ | 330 µH |
| $L_o$ | 10 µH |
| $C_1$ & $C_2$ | 100 µF |
| $C_{B1}$ & $C_{B2}$ | 10 µF |

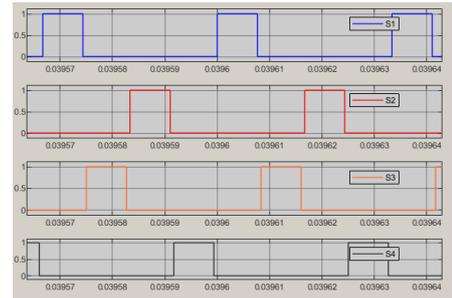

(a) Waveforms of switches signal

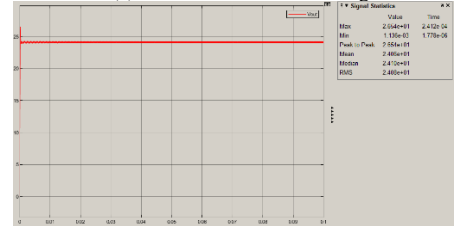

(b) Waveform of output voltage

Fig. 4. Waveforms of simulation

Fig. 4 (a) shows the bucked output voltage in MATLAB/Simulink simulation. It is obvious that with duty cycle equal to 0.235, the output voltage can be reduced to about 24V. According to the simulation, not only the voltage stresses of switches are equal to one-fourth of input voltage, but also the voltage stresses across the freewheeling diodes. This means that lower losses can be achieved in this converter.

For hardware experiment, the PCB of the selected structure is designed. In order to generate four PWM signals as Fig. 4 (b), the configuration of PWM blocks should be adjusted first. The four switches are turned on one by one and have a 1.5% OFF between the ON states of two PWM signals. The "AND logic" result of two PWM signals can become a delayed signal for $S_3$. By doing this, the four switching signals can be generated.

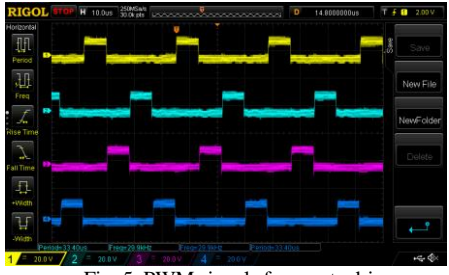
Fig. 5. PWM signals from gate drivers

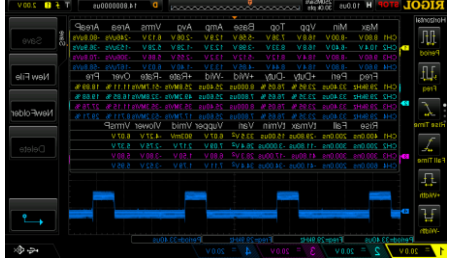
Fig. 6. Parameters of PWM signals

By applying these four PWM signals and then measure the input power to converter, the experimental results are collected, which are in agreement with the simulation results. Table 2 shows the results of output voltage in hardware experiment.

Table 2. Results of hardware experiment

| Input Voltage | Output Voltage | Simulation Results |
|---|---|---|
| 8 V | 65.4 mV | 54.88 mV |
| 9 V | 77.6 mV | 70.31 mV |
| 10 V | 99.4 mV | 96.3 mV |
| 12 V | 168.5 mV | 121.4 mV |

Until now, both simulation and hardware experimental results show that the selected converter can convert high voltage into low voltage with large duty cycle. It also shows that the converter has the advantages including lower voltage stress for switches, continuous input current etc. This ultra buck converter can be applied to high voltage conversion application.

### C. Analysis of the Results

Through simulation and experiment, this topology can obtain several advantages as above. However, it still has some drawbacks need to be addressed. The proposed topology can only operate for duty cycle less than 50%. If the duty cycles of four switches are set to larger than 50%, the output voltage will not follow the formula previously derived which means the converter cannot work properly. For example, if the duty cycles are set as 70%, and now the output voltage of circuit simulation is 122V.

From the analysis and experiment above, the voltage conversion ratio is improved a lot, from D to D/(4-D). However, if the converter is applied for converting 800V to 12V, the duty cycle should be set as 6%. Althoug this is much larger than the duty cycle ratio1.5% required for a conventional buck converter, it might still lead to higher loss or higher required control accuracy. Thus, the conversion ratio needs to be further improved. Furthermore, under the condition of 800-V-to12V, the input current becomes discontinuous.

From the above analysis, this topology does not apply any control technology to make the converter work under controlled conditions. In practical applications, control technology is necessary. The lack of control of the converter may lead to unstable output for practical applications, during the transient behaviour, or when the system jumps from one operating point to another. The power loss of switches is reduced a lot, due to the low voltage stress of switches in this topology. However, there are still high switching losses in actual operation. The non-zero current or voltage of the switch leads to nontrivial switching losses in the converter. A possible improvement is to apply soft switching that can eliminate switching losses by making the switching voltage or current zero.

### III. IMPROVEMENT ON SELECTED TOPOLOGY

#### A. Eight-phase topology

The topology proposed in [20], can be extended to more phases interleaved structure ideally to get higher voltage conversion ratio. The eight-phase extended structure of the four-phase converter above is deduced.

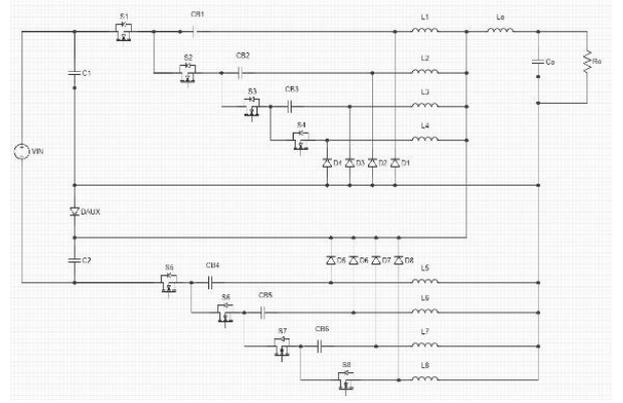
Fig. 7. Circuit of eight-phase topology

This eight-phase interleaved converter consists of eight switches $S_1$-$S_8$, eight freewheeling diodes, four blocking capacitors, two input capacitors, eight filter inductors $L_1$-$L_8$, one output filter inductor and one output filter capacitor. The proposed eight-phase topology has sixteen operating modes.

Mode 1: In this mode, $S_1$ is on and all diodes are conducting expect $D_1$. At this time, all other switches are not working. $L_1$ is charging through two paths, while other seven inductors are discharging. By using KVL law, several equations can be derived from the paths above:

$$V_{L1}(t) = V_{in} - V_{CB1} - V_{C2} = V_{C1} - V_{CB1} - V_{Lo}(t) - V_o \quad (14)$$
$$V_{Lo}(t) = V_{C1} + V_{C2} - V_{in} - V_o \quad (15)$$
$$V_{L2}(t) = V_{L3}(t) = \cdots = V_{L7}(t) = V_{L8}(t) = -V_o - V_{Lo}(t) \quad (16)$$
$$I_{Lo}(t) = I_{L1}(t) + \cdots + I_{L8}(t) - I_{in}(t) \quad (17)$$

Mode 2: In this mode, $S_1$ is turned off and all switches are off. Due to the inductor current cannot become zero immediately, the current in inductor $L_1$ still flows in two paths. All diodes are conducting and all inductors ($L_1$-$L_8$) are discharging. By using KVL law, several equations can be derived:

$$V_{L1}(t) = V_{L2} = \cdots = V_{L8} = -V_o - V_{Lo}(t) = V_{in} - V_{C1} - V_{C2} \quad (18)$$
$$V_{Lo}(t) = V_{C1} + V_{C2} - V_{in} - V_o \quad (19)$$

Mode 3: This mode begins as $S_5$ is turned on. $D_5$ is off with all other diodes are conducting. $L_5$ is charging through two paths. KVL law equations in this interval are as following:

$$V_{L5}(t) = V_{in} - V_{C1} - V_{CB4} = V_{C2} - V_{CB4} - V_{Lo}(t) - V_o \quad (20)$$
$$V_{Lo}(t) = V_{C1} + V_{C2} - V_{in} - V_o \quad (21)$$
$$V_{L1}(t) = V_{L2}(t) \cdots = V_{L8}(t) = -V_o - V_{Lo}(t) \quad (22)$$

$$I_{Lo}(t) = I_{L1}(t) + \cdots + I_{L7}(t) + I_{L8}(t) - I_{in}(t) \quad (23)$$

The next thirteen working modes are similar, with only one switch closed in each interval. With addition of four more blocking capacitors, the voltage stresses of many switches can be reduced. Same to the four-phase topology, in the actual work process, these eight switches in the converter may cause 256 different working states.

1. Converter Analysis

Apply the volt-second-balance (VSB) for the output inductor $L_o$.

$$\int_T V_{Lo}(t) = 0 \quad (24)$$

It can be concluded that $V_{Lo}(t)$ is zero. From the voltage-current equation of inductors, it can be concluded that the output current ripple $\frac{dI_{Lo}(t)}{dt}$ is regarded as zero.

Apply VSB for all inductors (L1-L8), equations can be derived as below :

$$(V_{C1} - V_{CB1} - V_o) \cdot D - V_o \cdot (1-D) = 0 \quad (25)$$
$$(V_{CB1} - V_{CB2} - V_o) \cdot D - V_o \cdot (1-D) = 0 \quad (26)$$
$$(V_{CB2} - V_{CB3} - V_o) \cdot D - V_o \cdot (1-D) = 0 \quad (27)$$
$$(V_{CB3} - V_o) \cdot D - V_o \cdot (1-D) = 0 \quad (28)$$
$$(V_{C2} - V_{CB4} - V_o) \cdot D - V_o \cdot (1-D) = 0 \quad (29)$$
$$(V_{CB4} - V_{CB5} - V_o) \cdot D - V_o \cdot (1-D) = 0 \quad (30)$$
$$(V_{CB5} - V_{CB6} - V_o) \cdot D - V_o \cdot (1-D) = 0 \quad (31)$$
$$(V_{CB6} - V_o) \cdot D - V_o \cdot (1-D) = 0 \quad (32)$$

Then the voltage conversion ratio can be concluded as

$$\frac{V_o}{V_{in}} = \frac{D}{8 - D} \quad (33)$$

As it can be observed, the voltage conversion ratio is improved approximately by a factor of two. In addition, voltage stress of switches and diodes can be reduced to one-eighth of input voltage. And the voltage stress of input capacitors can be derived as following:

$$V_{C1} = V_{C2} = \frac{2 \cdot V_{in}}{8 - D} \quad (34)$$

$$V_{CB3} = V_{CB6} = \frac{V_{in}}{4 \cdot (8 - D)} \quad (35)$$

$$V_{CB2} = V_{CB5} = \frac{V_{in}}{8 - D} \quad (36)$$

$$V_{CB3} = V_{CB6} = \frac{3 \cdot V_{in}}{2 \cdot (8 - D)} \quad (37)$$

A Similar analysis can be applied for this eight-phase topology, and an extended eight-phase converter can still have the same advantages as four phase converter above, including equal current sharing, continuous input current etc. Auxiliary diode is also added for unwanted resonance prevention.

2. Simulation

To validate the analysis, the simulation for extended phase converter is done. A prototype of proposed converter with 400V input voltage, 12V output voltage, 144W output power operating with an extended duty cycle of 0.24 for all switches is set up to demonstrate its functionality. All parameters of the Simulink circuit are shown in table.

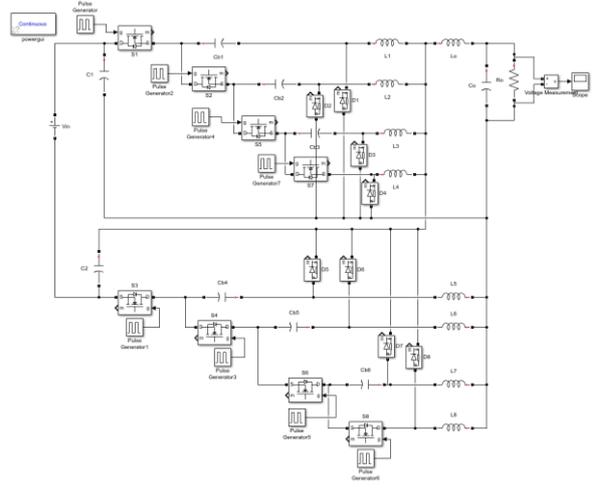

Fig. 8. Simulation circuit in MATLAB Simulink

The waveform of output voltage is shown in Fig. 9. It is very clear that the output voltage can be reduced to approximately 12V, with duty cycle equal to 0.24.

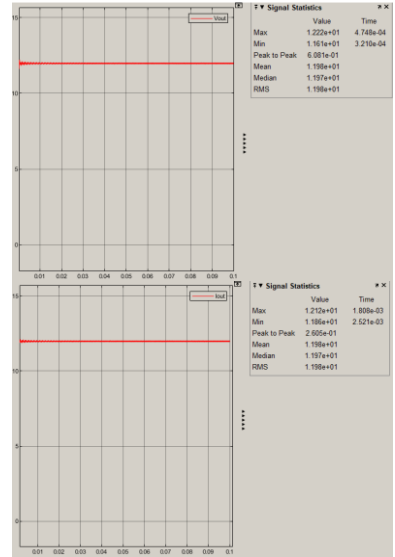

Fig. 9. (a) Waveform of output voltage $V_{out}$
(b) Waveform of output current $I_{out}$

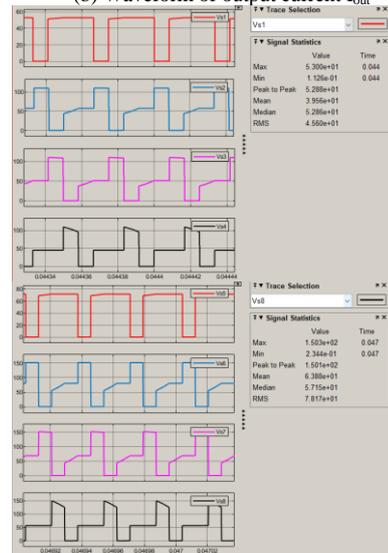

Fig. 10. (a) Voltage stresses of $S_1$-$S_4$
(b) Voltage stresses of $S_5$-$S_8$

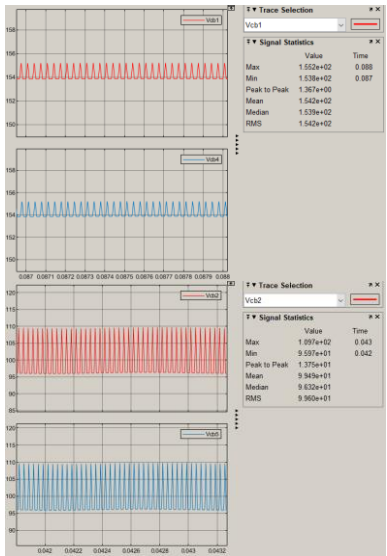

Fig. 11. (a) Voltage stresses of $C_{B1}$, $C_{B3}$
(b) Voltage stresses of $C_{B2}$, $C_{B5}$

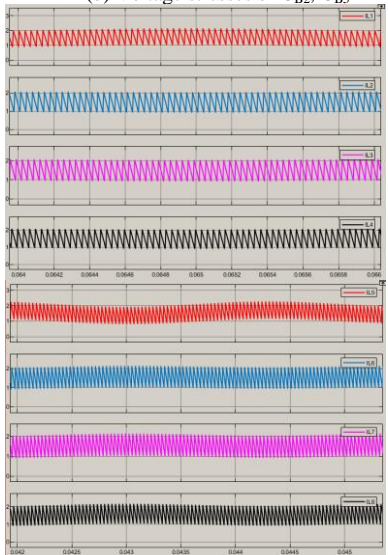

Fig. 12. Equal current sharing in eight phases ($I_{L1}$-$I_{L8}$)

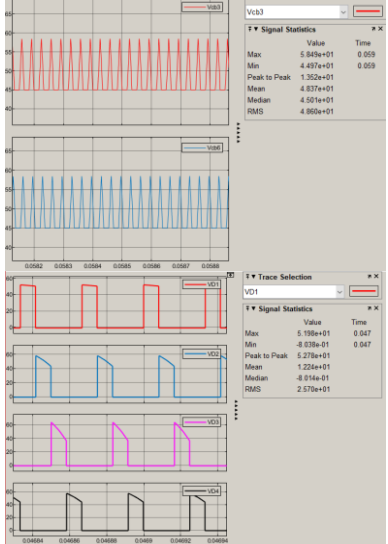

Fig. 13. (a) Voltage stresses of $C_{B3}$, $C_{B6}$
(b) Voltage stresses of diodes ($D_1$-$D_4$)

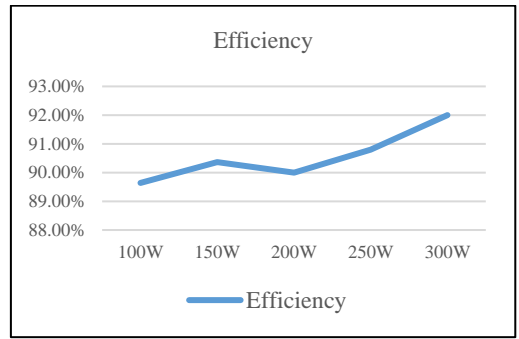

Fig. 14. Efficiency of the proposed eight-phase converter

Compared with the four-phase converter, the duty cycle is improved from 11.6% to 24%. It can prove that the voltage conversion ratio is indeed improved. Fig. 10 shows the voltage stress of the switches ($S_1$-$S_4$) and the diodes ($D_1$-$D_4$). It can be observed that $V_{S1}$ is 52.41V, which is only one-eighth of the input voltage. The voltage of the other three switches ($S_2$-$S_4$) is close to one-fourth of the input voltage, 110V. However, for fifth to eighth switches ($S_5$-$S_8$), these four switches suffer higher voltage stress than four switches above.

Same with [20], two input capacitors still withstand half of the input voltage. Fig. 12 shows equal current sharing can still be achieved in eight phases. From the new voltage conversion function, if 800V input voltage is to be converted to 12V output voltage, the duty ratio needs to be adjusted to 13%. Furthermore, according to Fig. 15, the input current can keep still continuous and the converter can still maintain 82%.

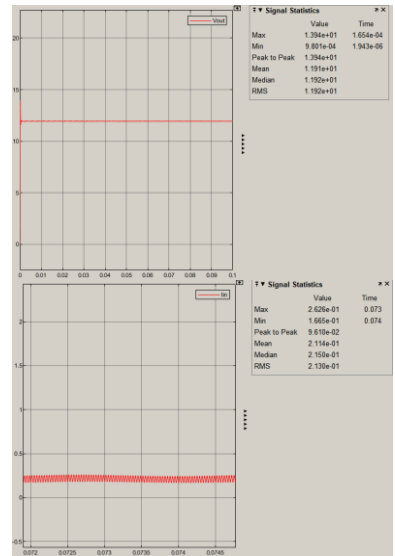

Fig. 15 (a) Waveform of output voltage;
(b) Waveform of output current

Using MATLAB simulation, the extended-phase converter can indeed have a higher conversion ratio, which can converter 800V to 12V with 13% duty cycle with efficiency of 82%. In addition, the advantages of the topology in [30] are still preserved, including low voltage stresses of the switches and diodes, continuous input current without any input filter etc.

At the same time, expanding to eight phases requires four more active switches and four more blocking capacitors to complete the operation, which causes the circuit more complicated. However, the improved eight-phase buck converter can be used for higher voltage conversion applications with better performances, such as in the application that converts 800V to 12V.

*B. ADC Control method*

Based on the converter to produce a constant 12V output voltage, the following two control methods are proposed.

The content of this method is to measure the value of the output voltage in real time by using the ADC component in PSoC development kit, and calculate the duty cycle required in real time by PSoC Creator. The calculated duty cycle is then applied to the PWM to adjust the switches operation states. Therefore, constant output voltage can be generated even though the input voltage changes.

The Delta Sigma Analog to Digital Converter (ADC_DelSig) is the front port of measuring instruments and it can be used in a variety of applications. Take a conventional buck converter which can generate constant 12V output voltage, as an example. Due to the conventional buck converter only requires an active switch, only one PWM block needs to be set up in the circuit. For the four-phase converter, the corresponding duty cycle adjustment is a bit more complicated. For example, the PWM signal of the second switch $S_2$ is generated by two PWM signals adding together. Therefore, each PWM blocks should make their own corresponding adjustments according to the calculated duty cycle. As shown infFig.17, the input of ADC block comes from Pin, which is connected to measure the input voltage. Through measuring the input supply, the ADC block will generate a hexadecimal feedback to system. The hexadecimal value can be converted to the corresponding float value by doing calculation, which represents the actual input voltage. And it can be used for calculating the required duty cycle.

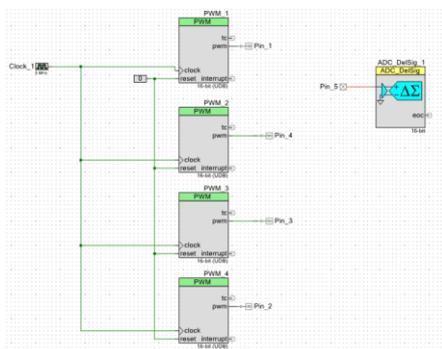

Fig. 17. Circuit in PSoC Creator

It is worth to mention that, due to the PSoC ADC component cannot afford high voltage, it is necessary to design a simple voltage divider circuit to reduce the voltage measured at the pin. Using voltage divider circuit, the ADC component measures an input voltage that is reduced by a factor of 101. In addition, in the algorithm code, this measured voltage value needs to be multiplied by 101 to represent the actual input voltage and then substituted into the equation.

By doing this, the actual input voltage can be measured by PSoC development board and as a feedback signal causes a change in duty cycle to get a constant output voltage. Therefore, a controlled buck converter can be achieved without any complicated components. However, there are actually many cases with special requirements or restrictions, and this control method still needs to be adjusted according to actual conditions. By combining the control method proposed above with the extended-phase buck converter, an ultrahigh voltage converter can be designed.

IV. CONCLUSION

This paper describes the development of DC/DC buck converters and the long-term challenges in the design of step-down power conversion system are investigated. Several converter models were chosen for simulation and the detailed analysis of one four-phase interleaved buck converter is shown. Hardware experiment is implemented on the selected topology. The strengths and weaknesses of three converters were also analyzed. This paper also proposes the extended-phase structure which is based on the four-phase topology. A control scheme using ADC components is proposed and verified by experiments. Based on the four-phase buck converter, by combining the proposed eight-phase converter with the ADC control method, an ultra buck DC-DC converter which can be applied for 800V-to-12V application is designed. In addition, the proposed converter can also obtain lower voltage stress for switches, continuous input current etc.